%% file: manuscript.tex
\documentclass{article}
\usepackage{arxiv}
\usepackage{authblk}
\usepackage[utf8]{inputenc} 
\usepackage[T1]{fontenc}    
\usepackage{hyperref}       
\usepackage{url}            
\usepackage{booktabs}       
\usepackage{amsfonts}       
\usepackage{nicefrac}      
\usepackage{microtype}      
\usepackage{fancyhdr}       
\usepackage{graphicx}       
\usepackage{float} 
\usepackage[T1]{fontenc}
\usepackage{dcolumn}
\usepackage{bm}
\usepackage{siunitx}
\usepackage{mathptmx}
\usepackage{etoolbox}
\usepackage{amsmath}
\usepackage{amssymb}
\usepackage{tabularx}
\usepackage{soul}
\setstcolor{red}
\DeclareSIUnit\angstrom{\text {Å}}
\usepackage{multirow}
\usepackage[style=chem-acs]{biblatex}
\title{Experimental Evidence of Quantum Drude Oscillator Behavior in Liquids Revealed with Probabilistic Iterative Boltzmann Inversion}

\author[1,2]{B. L. Shanks}
\author[2,3]{H. W. Sullivan}
\author[1]{P. Jungwirth}
\author[2]{M. P. Hoepfner}

\affil[1]{Institute of Organic Chemistry and Biochemistry of the Czech Academy of Sciences, Prague, CZH}
\affil[2]{University of Utah, \textit{Department of Chemical Engineering}, Salt Lake City, UT}
\affil[3]{University of Minnesota, \textit{Department of Chemical Engineering and Materials Science}, Minneapolis, MN}

\begin{document}

\maketitle

\begin{abstract}

The first experimental evidence of quantum Drude oscillator behavior in liquids is discovered using probabilistic machine learning-augmented iterative Boltzmann inversion applied to noble gas radial distribution functions. Furthermore, classical force fields for noble gases are shown to be reduced to a single parameter through simple empirical relations linked to atomic dipole polarizability. These findings highlight how neutron scattering data can inspire innovative force field design and offer insight into interatomic forces to advance molecular simulations.

\end{abstract}

% keywords can be removed

\newpage

\section{Introduction}

Atomic organization, structure, and self-assembly are foundational concepts in understanding and modeling the behavior of liquids at the atomic scale \cite{ornstein_accidental_1914,kirkwood_statistical_1951,hansen_theory_2013}. Solvent structure is known to have a large impact on the behavior of complex liquid phase processes, ranging from protein folding dynamics \cite{florova_explicit_2010, smith_force-field_2015, anandakrishnan_why_2019}, formation of lipid micelles in water \cite{brodskaya_effect_2015,deshmukh_water_2016}, to catalysis \cite{li_hydrogen_2022} and material separations \cite{kingsbury_microstructure_2018}. Consequently, the need for molecular models to accurately model liquid structure has increased the significance of obtaining precise experimental measurements to serve as benchmarks or optimization targets, ideally through the gold-standard techniques of wide-angle X-ray and neutron scattering. Scattering experiments probe interatomic distances and interactions by detecting individual scattering events over time, allowing for the sampling of atomic positions within the system at the same length scale as interatomic forces. This capability makes scattering data extremely valuable for the training and refinement of force fields (FFs) \cite{shanks_accelerated_2024}.

Within a classical liquid state theory paradigm, learning interatomic forces from experimental scattering data directly - the inverse problem in statistical mechanics - is a promising approach to enforce structural behavior in a FF model. Of existing inverse problem methods \cite{levesque_pair_1985,toth_determination_2001,toth_interactions_2007,ornstein_accidental_1914,percus_analysis_1958,wang_use_2010,agoodall_data-driven_2021,mullinax_generalized_2009,dunn_bottom-up_2016,delyser_bottom-up_2020,lyubartsev_calculation_1995,toth_iterative_2003,shanks_accelerated_2024}, iterative Boltzmann inversion (IBI) is one of the most widely used and easy-to-implement techniques. In a nutshell, IBI refines pair potentials based on their quality-of-fit to a target set of site-site partial radial distribution functions, $g_{\alpha,\beta}(r)$ (e.g., the O-O correlation in water). The algorithm proceeds iteratively until the target structure is reproduced to a pre-specified degree of accuracy.

Although experiencing a renewed interest for training coarse-grained FFs, IBI has its roots in a study on the uniqueness of the pair interaction potential in relation to the pair correlation functions from Henderson \cite{henderson_uniqueness_1974}. Based on this now dubbed "Henderson's inverse theorem", Wolfram Schommers designed a variational method to extract pair potentials for his studies of liquid gallium \cite{schommers_pair_1983}, which would later evolve into the numerical IBI type methods existing today. Applications of Schommer's algorithm range from neutron scattering analysis with empirical potential structure refinement \cite{soper_empirical_1996}, coarse-grained FF design \cite{moore_derivation_2014,wang_development_2020}, to correcting structural predictions in machine learning potentials \cite{matin_machine_2024}. Recently, more rigorous mathematical formulations of the problem have been explored to try and elucidate fundamental properties of IBI within the framework of functional analysis \cite{hanke_well-posedness_2018,frommer_note_2019,frommer_variational_2022}. 

Of course, IBI has well-documented limitations that restrict its transferability to complex model systems and experiments. For instance, it is generally observed that IBI potentials depend on the chemical environment and thermodynamic state of the training data \cite{,rosenberger_comparison_2016}, prompting the general shift towards multi-objective optimization over a range of thermodynamic state points \cite{moore_derivation_2014}. In the context of the theory from which these algorithms arise, this state-dependence is not surprising. Explicitly, Henderson's inverse theorem only guarantees uniqueness of the pair potential for isotropic fluids with pairwise additive interactions at fixed density and pressure \cite{frommer_note_2019}. The former condition is violated for pairwise additive two component mixtures, while both conditions are generally violated for molecular liquids. Another problem is that, unlike atomistic simulations in which the site-site partial radial distribution functions can be computed directly from a trajectory, the experimental equivalent is not computable uniquely \cite{soper_uniqueness_2007}. The fact that applying IBI to experimental data already violates assumptions of the theory, and that the target experimental data is potentially an inaccurate representation of the true atomic structure, leaves us in a difficult state-of-affairs for learning potentials in a rigorous and reproducible way.

However, it is now possible to mitigate many of the above challenges with the use of machine learning \cite{shanks_transferable_2022}. Specifically, we recently proposed a probabilistic IBI algorithm, i.e., structure optimized potential refinement (SOPR), to extract transferable forces from experimentally derived RDFs for fluid phase systems. The key advance that SOPR provides is a Gaussian process regression (GPR) step on the refinement equation aimed at mitigating numerical instability and overfitting to uncertain experimental data. GPR enforces continuity, differentiability, and can suppress spurious long-range potential oscillations. Conceptually, GPR takes the standard IBI algorithm and embeds rigorous uncertainty quantification \cite{rasmussen_gaussian_2006} into its interatomic potential prediction. Application of SOPR to neutron scattering data of noble gases (Ne, Ar, Kr, and Xe) has reproduced both structural and thermophysical properties along the vapor-liquid line with accuracy that matches or exceeds state-of-the-art classical force fields \cite{shanks_transferable_2022}.

The ability to compute accurate interaction potentials directly from scattering measurements opens the door to deeper insights into interatomic forces in liquids. SOPR potentials, serving as a two-body mean-field approximation of the quantum many-body interactions that govern liquid behavior, overcome the system size limitations of quantum models while remaining fully nonparametric and flexible. The Bayesian nature of the SOPR method also reduces model selection bias and parametric uncertainty, offering an experimentally validated method to quantify deviations from empirical functional forms like Lennard-Jones and to assess the magnitude of many-body effects against \textit{ab initio} dimer potentials. Moreover, analyzing SOPR potentials across thermodynamic states or atomic numbers can reveal valuable links between fundamental atomic properties and classical representations of interatomic forces \cite{shanks_uncertainty-aware_2024}.

The analysis of noble gases provides a compelling test case for this approach. Noble gases largely satisfy the assumptions of the Henderson inverse theorem (excluding pairwise additivity), and their long-standing role in studies of intermolecular interactions makes them ideal candidates for examining consistencies between classical and quantum models of interatomic forces. To this aim, we deconstructed noble gas SOPR potentials using classical parameterizations of atomic size and dispersion energy and compared them to the atomic dipole polarizability. The atomic dipole polarizability was selected as the variable quantity because it known to be closely related to atomic size in quantum models of electron polarization \cite{fedorov_quantum-mechanical_2018} and is present in the leading term of the dispersion energy multipole expansion \cite{london_general_1937}. Empirical investigation of the atomic size derived from SOPR potentials revealed that both collision diameter (the $\sigma$ parameter in classical force fields) and hard-particle diameter exhibit power law scaling that aligns precisely with the quantum Drude oscillator model of electronic polarization. This result, to our knowledge, is the first observation of quantum Drude oscillator type behavior in the liquid state.

The empirically derived scaling laws were subsequently used to predict force field parameters for noble gases outside the training set, specifically helium (He), radon (Rn), and oganesson (Og). These predictions matched independently optimized force fields up to the estimated uncertainty in the pair potential reconstruction (excluding Og, for which no experimental data of liquid phase properties currently exists). Notably, the excellent quantitative agreement with a Rn force field optimized to critical point data \cite{mick_prediction_2016} is striking. This suggests that SOPR can be leveraged to predict accurate force field parameters that are fully consistent with quantum representations of atomic dipole polarizability, despite being derived within a purely classical framework. Moreover, this result provides a clear example of dimensionality reduction in force field design, where the problem of developing classical potentials can be reduced to computing a single parameter and applying simple scaling relations learned from experimental structure data. These results demonstrate that nonparametric potentials derived from neutron scattering patterns not only align with quantum mechanical theory but also have the potential to guide both classical and next-generation quantum mechanical force field development in subtle yet powerful ways. 

\section{Theory and Computational Methods}

In practice the pairwise additive approximation has been widely successful at modeling thermodynamic and dynamic properties of liquid state systems at a comparatively low computational cost. Under the additional assumption that the liquid is isotropic, the two-body effective potential is uniquely defined by the radial distribution function according to Henderson's inverse theorem \cite{henderson_uniqueness_1974}. This theorem can be shown to be mathematically equivalent to the condition that for any point $r_i \in \mathbb{R}^+_0$, the product of the difference in the effective pair potential, $\Delta v_2(r_i)$, and radial distribution function, $\Delta g(r_i)$, must be non-positive,

\begin{equation}\label{eq:henderson}
    \Delta v_2(r_i)\Delta g(r_i) \leq 0
\end{equation}

\noindent establishing a variational framework to refine a reference potential to experimentally derived radial distribution functions. The two-body potential derived from the experimental scattering data is therefore the unique potential that gives an equality in eq \ref{eq:henderson}.  

\subsection{Structure Optimized Potential Refinement}

\begin{figure}
    \centering    \includegraphics[width=0.9\linewidth]{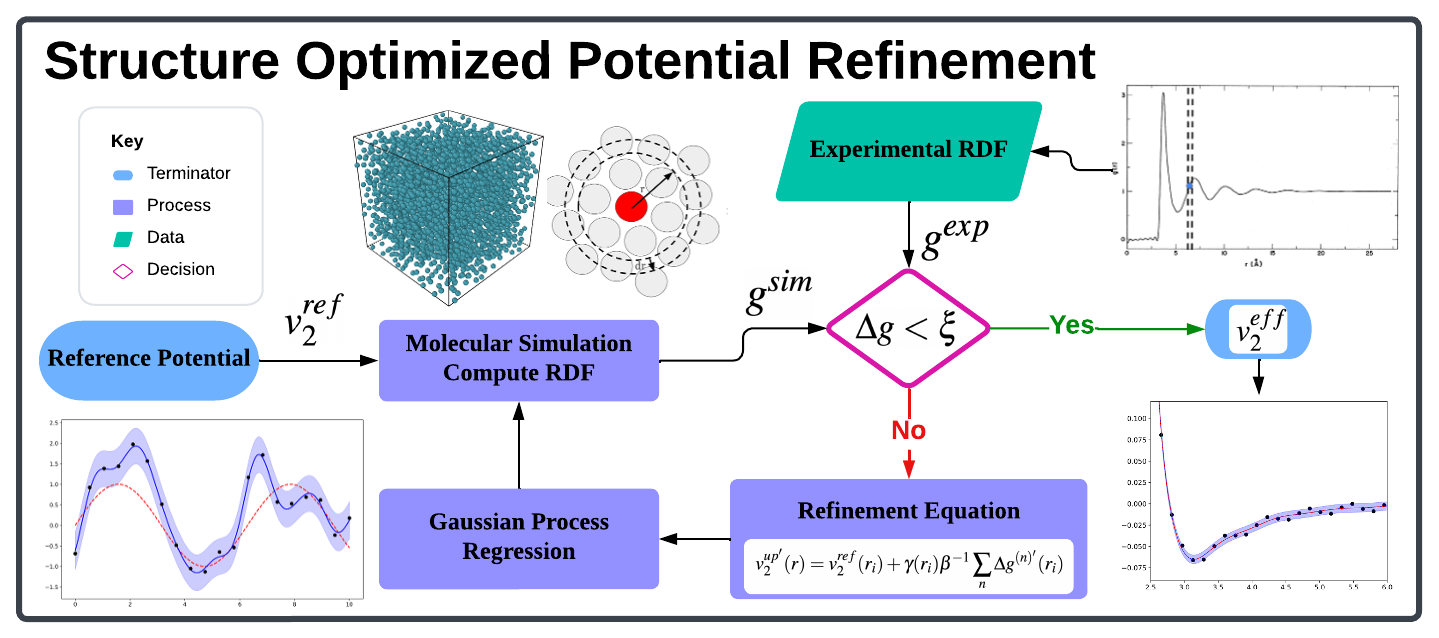}
    \caption{General overview of the structure optimized potential refinement algorithm. The loop between the molecular simulation, potential refinement equation, and GPR is always convergent to a unique potential for pairwise additive and isotropic systems according to Henderson's inverse theorem.}
    \label{fig:sopr}
\end{figure}

Structure optimized potential refinement (SOPR) is a probabilistic iterative Boltzmann inversion algorithm designed to learn effective two-body potentials from neutron/X-ray scattering derived radial distribution functions \cite{shanks_transferable_2022}. Whereas most IBI methods are used for coarse-grained simulations \cite{moore_derivation_2014}, SOPR is uniquely designed to handle experimental data. In a nutshell, SOPR takes as inputs a reference pair potential, $v_2^{ref}$, and experimentally derived radial distribution function, $g^{exp}$, and computes an effective pair potential, $v_2^{eff}$, that reproduces the given experimental data (Figure \ref{fig:sopr}). This computation is performed via iterative potential refinement, in which (1) a molecular simulation using the given reference potential is used to produce a simulated radial distribution function, $g^{sim}$, (2) the difference between the simulated and experimental radial distributions functions is calculated, $\Delta g = g^{sim} - g^{exp}$, and (3) the pointwise Henderson inverse theorem is applied to estimate an updated potential, $v_2^{up'}$, for the next iteration via a refinement equation,

\begin{equation}\label{eq:refinement}
    v_{2}^{up'}(r) = v_2^{ref}(r_i) +  \gamma (r_i) \beta^{-1} \sum_n \Delta g^{(n)'} (r_i)
\end{equation}

\noindent where $i$ is the radial index, $n$ is the iteration number, $\beta$ is the inverse thermal energy $(k_B T)^{-1}$, and $\gamma(r_i) > 0$ is an empirical scaling function used to dampen the potential correction. Note that eq \eqref{eq:refinement} is just one refinement equation example that satisfies eq \eqref{eq:henderson} and modifications to this function can lead to changes in convergence speed and stability. 

At this stage, traditional IBI repeats steps (1)-(3) until the simulated and experimental radial distribution functions converge to each other ($\Delta g < \xi$ $\forall r$). $\xi \in \mathbb{R}^+$ is a tolerance limit and should be approximately equal to the noise in the target RDF. However, SOPR introduces an additional machine learning step (4) Gaussian process regression (GPR) on the updated potential. Conceptually, we are assuming that the updated potential is a \textit{sample} from a distribution of possible pair potential functions that are approximately distributed as an infinite dimensional multivariate Gaussian over its input parameters,

\begin{equation}\label{eq:GPR}
    v^{up'}_{2} \sim \mathcal{N}(\mu(\mathbf{r}), \mathbf{K}(r_i,r_j))
\end{equation}

\noindent where $r_i, r_j$ are any pair of radial positions, $\mu(\mathbf{r})$ is a mean function, and $\mathbf{K}(r_i,r_j)$ is a covariance function (or kernel) describing the relatedness of observations $v^{up'}_{2}(r_i)$ on $v^{up'}_{2}(r_j)$ \cite{rasmussen_gaussian_2006}. The specification of the mean and kernel functions are where we can enforce physics based knowledge such as continuity and differentiability (i.e., smoothness) and long-range tail behavior ($\lim_{r \to \infty}v(r) = 0$). It is then a straightforward matrix calculation to compute the most probable potential given the iterative potential refinement estimate as,

\begin{equation}\label{eq:gp}
    v^{up}_{2}(r') = \mathbf{K}^T_{r',r}[\mathbf{K}_{r,r} + \sigma_{noise}^2\mathbf{I}]^{-1}v^{up'}_{2}(r)
\end{equation}

\noindent where $v^{up}_2(r')$ is the estimate for the structure-optimized potential at iteration $n$, $\sigma_{noise}^2$ is the variance in the function resulting from random noise, and $\mathbf{K}_{r',r}$ is shorthand notation for the kernel function. Generally, the $\sigma_{noise}^2$ hyperparameter can be inferred using hierarchical Bayesian inference for each regression step, but for our purposes it was sufficient to choose a value of 0.01 to capture known noise levels observed during refinement. The GPR step has been shown to reduce numerical instability and enforce physically justified behavior in the refined potentials, allowing SOPR to address well-known problems with IBI such as artifacts associated with intermediate and long-range structural correlations and potential non-uniqueness in heterogeneous systems (i.e., molecular liquids and mixtures).

In this work, SOPR was applied to neutron scattering data for four noble gas species (Ne, Ar, Kr, Xe) \cite{yarnell_structure_1973,bellissent-funel_neutron_1992,barocchi_neutron_1993} to learn nonparametric potentials for subsequent analysis. The methodology closely resembles that from our previous work \cite{shanks_transferable_2022} but includes updated simulation details for HOOMD-Blue version 4.7.0 \cite{anderson_hoomd-blue_2020}. Details of the molecular simulation set-up, SOPR parameters (reference potentials, scaling function, convergence tolerance, etc), mean and kernel selection for GP regression, and validation of the resulting potentials can be found in the Appendix.

\subsection{Short Range Interactions: Atom Size Estimation}

In classical force field design, the atomic size parameter, often the "collision diameter" or $\sigma$, is used in empirical potentials like the (12-6) Lennard-Jones (LJ) and $(\lambda-6)$ Mie potentials. The collision diameter represents the repulsive core of the atom, where Pauli exclusion causes strong repulsion. $\sigma$ is typically defined as the diameter where the pair potential shifts from positive to negative at short range. Interestingly, $\sigma$ can vary significantly across different FFs (c.f. Madrid 2019 \cite{zeron_force_2019} and CHARMM \cite{mackerell_jr_empirical_2004}), which can easily result in a poor quality of fit to experimentally derived radial distribution functions. For instance, it has been estimated that the uncertainty in the $\sigma$ parameter is less than 0.1 $\si\angstrom{}$ when using structure factor measurements as an optimization target \cite{shanks_bayesian_2024}. Consequently, two classical FFs with $\sigma$ values differing by more than 0.1 $\si\angstrom{}$ are likely to produce significantly different structure factor predictions, with at least one diverging notably from experiment. 

The van der Waal (vdW) diameter is another important quantity in quantum mechanical calculations that often appears in density functional theory (DFT) \cite{wu_empirical_2002}. Specifically, the vdW diameter is directly influential to the so-called "vdW force" present in DFT functionals \cite{berland_van_2015}. The standard definition for the vdW diameter used today is the distance from the center of an atom at which Pauli exclusion and London dispersion forces are balanced. In classical pair potential models, this point occurs at the diameter where the force is zero, typically located at the bottom of the potential well. In DFT, the exchange-correlation energy functional, which describes interactions of many-particles in the standard Kohn-Sham DFT method, includes vdW forces in an approximate manner \cite{berland_van_2015}. By directly computing nonparametric classical potentials from experimental data, we can compare vdW diameter estimates in many-body liquid-phase systems with those obtained from quantum mechanical calculations. While the idea to learn vdW diameters from scattering data dates back to the work of Bondi \cite{bondi_van_1964}, to our knowledge, no attempts have been made so far to estimate them using nonparametric potentials derived from scattering. 

\begin{figure}
    \centering
    \includegraphics[width=0.9\linewidth]{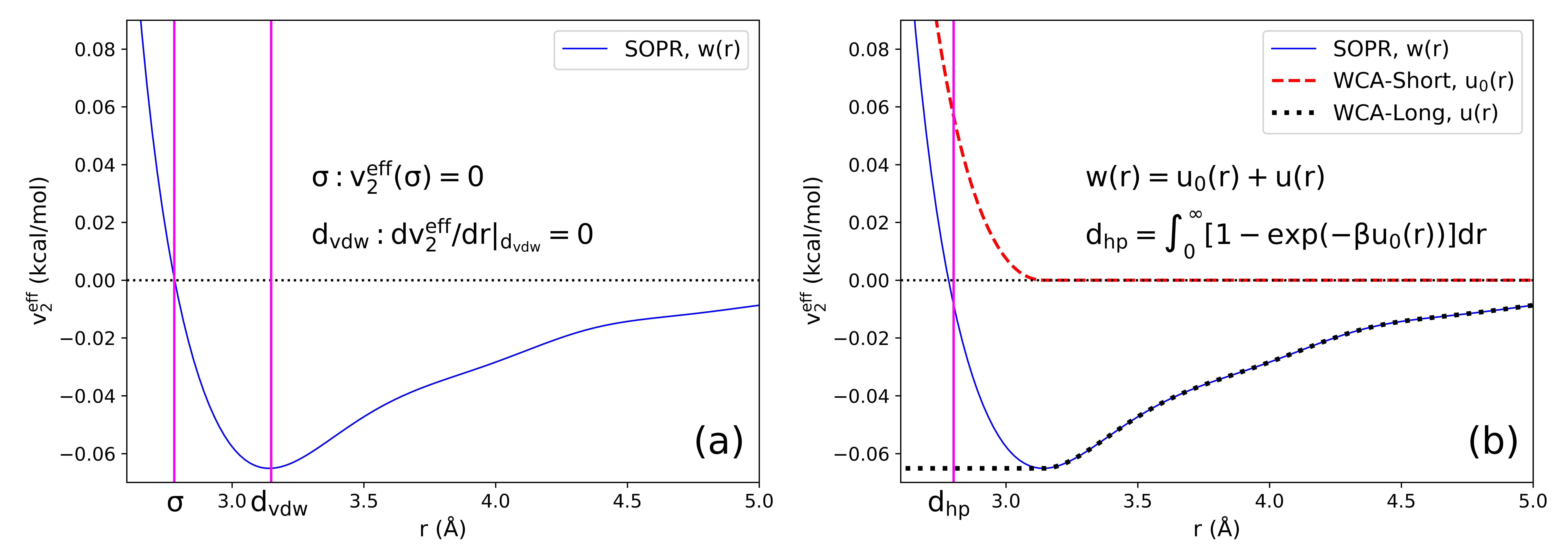}
    \caption{WCA separation of the SOPR potential into a short-range repulsive reference potential and long-range perturbation.}
    \label{fig:WCA}
\end{figure}

Finally, for model developers interested in entropy driven self-assembly \cite{glotzer_anisotropy_2007}, the effective hard-particle diameter defines particle size. Here the hard-particle diameter was determined using the Weeks-Chandler-Andersen (WCA) perturbation theory \cite{weeks_role_1971}, in which the total potential, $w(r)$, is separated into two parts: a short-range repulsive reference, $u_0(r)$ and a long-range attractive perturbation $u(r)$,

\begin{equation}\label{eq:WCA}
    w(r) = u_0(r) + u(r).
\end{equation}

\noindent such that,

\begin{equation}
u_0(r) = 
\begin{cases} 
w(r) + \epsilon & \text{if } r < r_{\epsilon},\\
0 & \text{otherwise}.
\end{cases}
\end{equation}

\begin{equation}
u(r) = 
\begin{cases} 
-\epsilon & \text{if } r < r_{\epsilon},\\
w(r) & \text{otherwise}.
\end{cases}
\end{equation}

\noindent where $\epsilon$ is the pair potential well depth and $r_\epsilon$ is the radial position corresponding to the potential energy $\epsilon$. Eq \eqref{eq:WCA} is shown graphically in Figure \ref{fig:WCA}(b) for the SOPR potential of liquid neon. 

Once the potentials are decomposed using the WCA separation, an equivalent hard-particle diameter, $d_{hp}$, can be estimated from the repulsive part of the potential \cite{barker_perturbation_1967,weeks_role_1971,hansen_theory_2013} according to the following equation,

\begin{equation}\label{eq:hardparticle}
    d_{hp} = \int_0^\infty [1 - \exp(-\beta u_0(r))] dr
\end{equation}

\noindent such that the integrand $[1 - \exp(-\beta u_0(r))]$ is unity at low $r$ since $\exp(-\beta u_0(r))$ is negligibly small due to the y-asymptotic behavior of $u_0(r)$. At longer range, $u_0(r)$ is zero by definition, which sends the integrand to zero. Note that this hard-particle diameter falls out of a second order approximation of the functional Taylor expansion of the reduced excess free energy density, $\phi = -\beta F^{ex} / V$, in powers of the blip function, $\Delta e(r)$, defined as the difference between Boltzmann factors of the short-range repulsive reference and a system of equivalent hard spheres,

\begin{equation}
    \Delta e(r) = \exp(-\beta u_0(r)) - \exp(-\beta u_d(r))
\end{equation}

\noindent where $u_d(r)$ is an equivalent hard-particle reference potential. A key strength for performing this analysis on SOPR potentials is that the soft-core reference system, empirical hard-particle size, and blip function can now be estimated directly from experiments, allowing for comparisons between experimentally inferred potentials and those derived from liquid state theory. 

The motivation for selecting these definitions of atomic size to describe short-range interactions are three-fold. First, the collision diameter, vdW diameter, and hard-particle diameter are common parameters across a wide range of computational models, covering classical FFs, DFT, and hard-particle Monte Carlo, respectively. In effect, any relationships derived from these three parameters will be widely applicable across numerous fields of computational molecular science. Second, we wanted to leverage the low computational cost of SOPR to see if certain size parameters follow known theoretical behaviors while others do not. Such an undertaking would be far more expensive for existing \textit{ab initio} methods due to the large number of atomic clusters required to represent a liquid phase. Finally, it is not known whether or not any trends exist in empirical atomic size parameters like the collision diameter and hard-particle diameter. Elucidating these relationships could provide an experimentally validated FF design strategy or be used as initial guess for optimization.

\subsection{Long Range Interactions: Attractive Forces}

The treatment of long-range interactions differs substantially between classical and quantum mechanical models of interatomic forces, complicating direct comparisons between SOPR potentials and quantum models of dispersion interactions. In this work, we approximate the classical dispersion energy as the minimum energy of the SOPR pair potential, equivalent to the $\epsilon$ parameter in the familiar Lennard-Jones model, and perform linear regression on a power law relation of the form, $\epsilon = a\alpha^{b}$, where $a,b \in \mathbb{R}$. As demonstrated later in the manuscript, even this simplified approach reproduces the $\epsilon$ parameter for Rn within the expected parametric uncertainty of structure-derived potentials.

\subsection{Atomic Dipole Polarizability and Quantum Drude Oscillators}

Atoms and molecules are electrically polarizable particles, and their degree of polarization is well-known to correlate with their effective size \cite{mandel_molecular_1958}. In condensed phases, where interatomic interactions significantly impact the static and dynamic behavior of the system, electronic polarization effects can become significant in response to the chemical environment. These effects are especially important in biomolecular systems where polarization is known to play a large role in interactions between ions and proteins, lipids, and charged amino acids. Therefore, there is currently a widespread effort to incorporate electronic polarization models in classical FFs using both explicit methods, such as fluctuating charge models or classical drude oscillators \cite{lamoureux_simple_2003}, and implicit mean-field electron continuum corrections (ECC) \cite{leontyev_polarizable_2012,kirby_charge_2019,cruces_chamorro_building_2024, nencini_effective_2024}. 

Of the explicit models of electron polarization, the quantum Drude oscillator (QDO) is emerging as a popular method to describe electronic polarization from first principles \cite{whitfield_low_2007,jones_electronically_2013,hermann_first-principles_2017}. In the QDO model, the interaction between the electrons and nucleus is represented as a harmonic term between a \textit{drudon} quasiparticle with charge $-q$ and a classical quasinucleus of charge $+q$. Each quasiparticle interacts with other particles in the system through Coulombic forces. In a spherically symmetric QDO, polarizability can be calculated from second-order perturbation theory by constructing a test charge perturbed Hamiltonian and solving for the induced multipole moments \cite{jones_quantum_2013}. Of particular interest to the biomolecular FF community is the construction of computationally efficient two-body interatomic potentials consistent with QDO models \cite{khabibrakhmanov_universal_2023}. 

Another benefit of the QDO model is its simple mathematical structure, which makes it easy to derive relationships between atomic properties \cite{tkatchenko_fine-structure_2021}. For example, Federov \textit{et al.} \cite{fedorov_quantum-mechanical_2018} showed that the vdW diameter is related to atomic dipole polarizability via a power law,

\begin{equation}\label{eq:qdoscaling}
    R_{vdW} \approx a \alpha^{1/7}
\end{equation}

\noindent where $\alpha$ is polarizability and $a \approx 2.54$ is a constant fit to target vdW diameter data. It has been argued that this relation, which touts reasonably accurate predictions for 72 elements, obviates the need to compute vdW diameters since the atomic dipole polarizability is easily measured experimentally with index of refraction or Rayleigh scattering probability methods \cite{hohm_interferometric_1990, minemoto_measuring_2011, gaiser_polarizability_2018} or computed \cite{mitroy_theory_2010,gobre_efficient_2016}. Note that this relation differs from the classical drude oscillator (CDO) solution, which scales with $R \propto \alpha^{1/3}$.

These scaling laws provide a convenient consistency check between SOPR potentials and quantum mechanical representations of electronic polarization. To this aim, we fit the QDO power law scaling in eq \eqref{eq:qdoscaling} to SOPR estimated vdW diameters, collision diameters, and hard-particle diameters to elucidate whether or not the structure-derived potentials are consistent with quantum mechanical behavior of electronic polarization. The resulting empirical QDO power relation was then used to estimate collision diameters for the noble gases He and Rn. 

\section{Results and Discussion}

Our methodology combines the following three steps: (1) calculation of the $\sigma$ parameter, vdW diameter, hard-particle diameter, and dispersion energy from noble gas SOPR potentials (2) linear regression of power law relations between these parameters and atomic dipole polarizability, and (3) application of the resulting relations to estimate $\sigma$ and $\epsilon$ parameters in noble gases outside of the training set. In a nutshell, the results show that the SOPR-QDO line derived solely from liquid noble gas scattering data is quantitatively accurate at predicting optimal $\sigma$ and $\epsilon$ parameters for classical models of noble gases.

\subsection{Analysis of SOPR Potentials and QDO Scaling Laws}

\begin{figure}[H]
    \centering \includegraphics[width=0.9\linewidth]{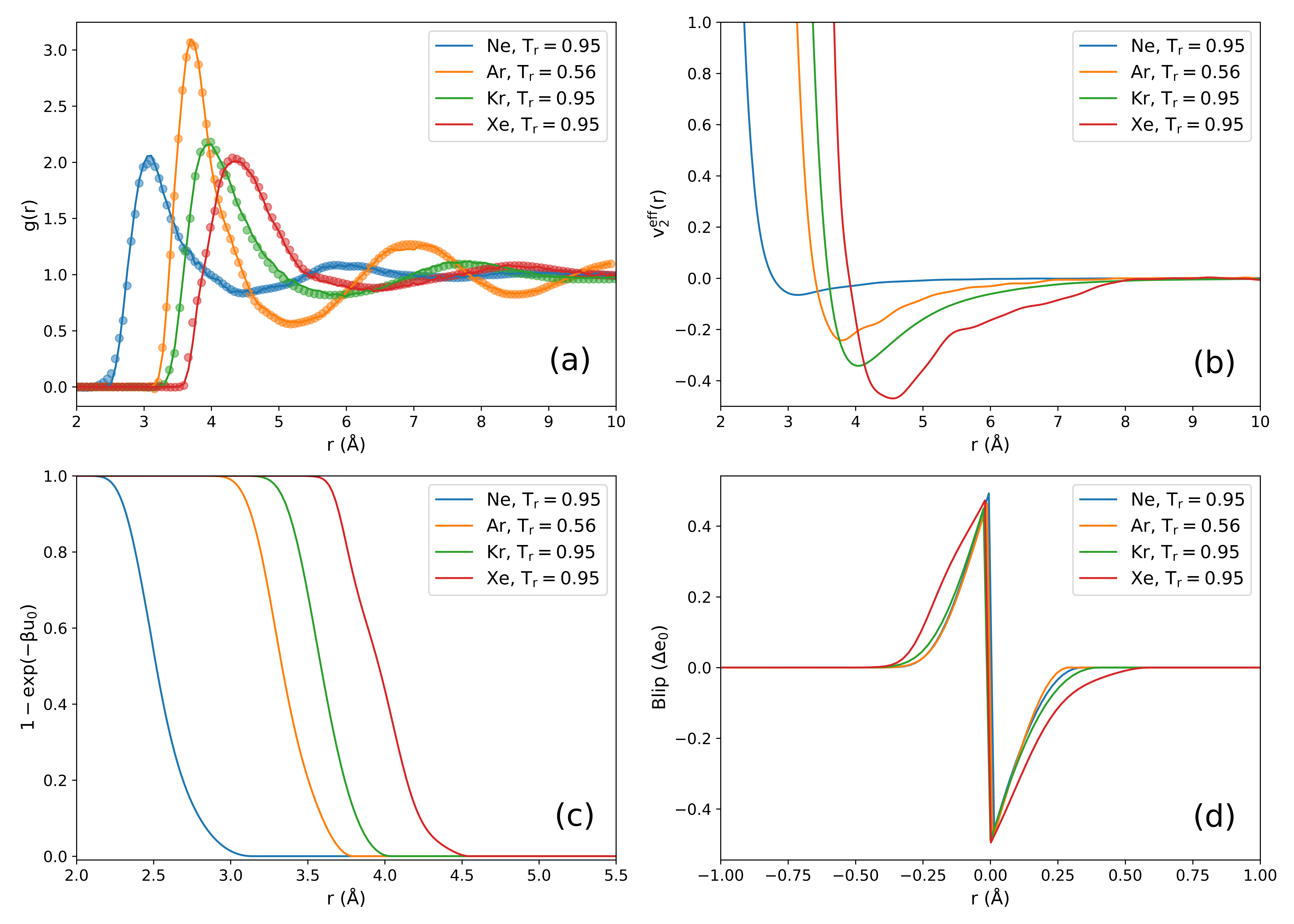}
    \caption{SOPR analysis for Ne, Ar, Kr, and Xe at a single state point. (a) Experimentally derived radial distribution functions (markers) compared to SOPR simulation RDFs (lines). (b) Converged SOPR potentials. (c) Integrand of the soft-core repulsive potential after after Weeks-Chandler-Andersen separation with the Barker-Henderson hard-particle diameter approximation. (d) Neutron scattering derived blip functions.}
    \label{fig:hp_integral}
\end{figure}

Radial distribution functions, SOPR potentials, hard-particle integrands, and blip functions calculated from neutron scattering data of Ne, Ar, Kr, and Xe are shown in Figure \ref{fig:hp_integral}. The excellent quality of fit to the experimental radial distributions is expected according to the Henderson inverse theorem. Also, the SOPR potentials themselves are well-behaved in comparison to other IBI potentials derived from experimental data \cite{schommers_pair_1983,soper_partial_2005}. Figure \ref{fig:hp_integral} (c) shows the integrand of the Barker-Henderson hard-particle equation after WCA separation of the SOPR potential and (d) the blip function. The widening of the blip function with atom identity indicates a larger deviation from hard-particle behavior, consistent with the progressively larger and more polarizable electron clouds of Ne, Ar, Kr, and Xe, respectively. Atomic vdW diameters, collision diameters, and hard-particle diameters computed from the SOPR potentials are provided in Table \ref{tab:diameters}.

\begin{table}[H]
\centering
\caption{\label{tab:diameters} Summary of atomic size and dispersion energy parameters determined from SOPR potentials. All lengths are given in units of $\si\angstrom{}$.
}
\begin{tabular}{l c c c c c c}
\hline
\textrm{Element}&
\textrm{$\sigma$}&
\textrm{vdW}&
\textrm{$d_{hp}$}&
\textrm{$\epsilon$ (kcal/mol)}&
\textrm{$\alpha$ ($\si\angstrom{}^3$) \cite{gobre_efficient_2016}}
\\
\hline
Ne  & 2.77 & 3.15 & 2.80 & 0.065 & 0.396 \\
Ar  & 3.40 & 3.79 & 3.51 & 0.242 & 1.661\\
Kr  & 3.58 & 4.05 & 3.64 & 0.342 & 2.528\\
Xe  & 3.91 & 4.55 & 3.96 & 0.469 & 4.118\\
\hline
\end{tabular}
\end{table}

\begin{figure}[H]
    \centering \includegraphics[width=0.9\linewidth]{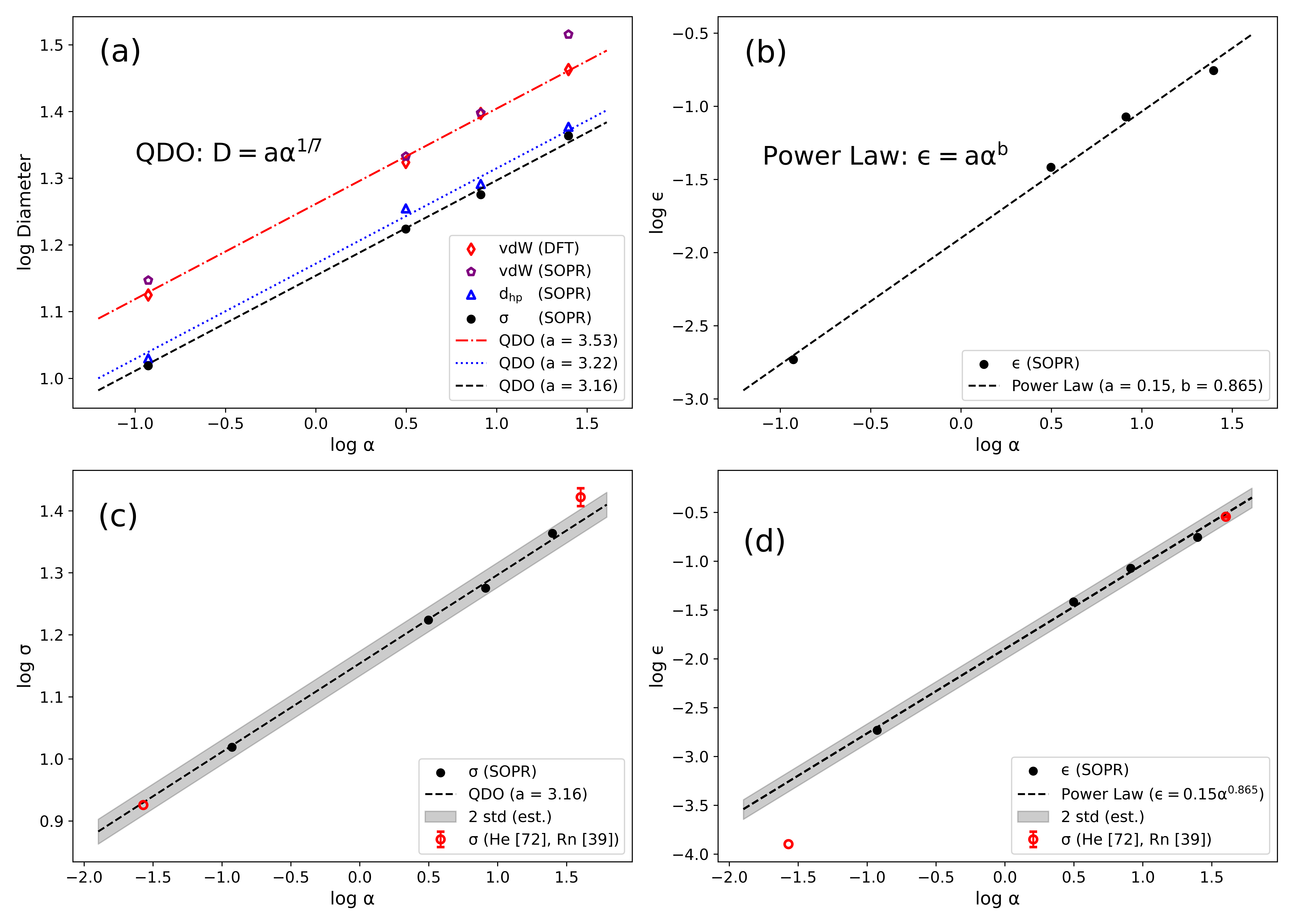}
    \caption{Interatomic potential parameters derived from SOPR compared to a QDO power law (for atomic size) and variable power law relation (for dispersion energy) in the noble gas series. (a) Empirical quantum Drude oscillator scaling law, $D \propto \alpha^{1/7}$, (dashed lines) compared to \textit{ab initio} calculated vdW diameter (red diamonds) and neutron scattering derived vdW diameter (violet pentagons), effective hard particle diameter (blue triangles) and interaction potential collision diameter (black circles). (b) Dispersion energy at the pair potential minimum vs variable power law relation. (c-d) Extrapolation of QDO and variable power laws to atomic dipole polarizabilities of He and Rn parameters with uncertainty estimation. Red markers represent independently optimized force field parameters.}
    \label{fig:polarize}
\end{figure}

A comparison between the SOPR potential atomic size predictions and atomic dipole polarizabilities is presented in Figure \ref{fig:polarize}(a) while dispersion energy is presented in Figure \ref{fig:polarize}(b). Power law constants, $a,b \in \mathbb{R}$, were computed using linear least-squares regression to the equation, $c = a \alpha^{b}$. For the atomic size predictions, the power law exponent, $b$, was fixed to the QDO value of $1/7$, whereas for the dispersion energy $b$ was treated as an unknown. The power law relations and fitting parameters are recorded in Figure \ref{fig:polarize} (a-b).

Collision and hard-particle diameters show an excellent linear fit in the log-log plot, with a power law exponent of 1/7, providing the first experimental evidence of QDO-like behavior experimental data in the liquid-state. In contrast, the SOPR vdW diameter deviates from the prediction by Fedorov \textit{et al.} \cite{fedorov_quantum-mechanical_2018} for liquid Xe, which is expected since the QDO relation, derived in vacuum, does not account for the many-body interactions in the dense polarizable liquid. This discrepancy underscores the limitations of naively applying QDO scaling to classical force fields: while short-range interactions such as collision and hard-particle diameters align with the QDO model, the vdW diameter, influenced by many-body effects, does not.

Furthermore, this deviation may reveal deeper complexities in QDO behavior under liquid-state conditions. One possible explanation is that the solution to the perturbed QDO Hamiltonian for many-body systems at liquid densities does not follow a power law in atomic dipole polarizability, unlike the symmetric single QDO with a test charge perturbation. In this case, the SOPR vdW data may suggest the presence of another term in the solution that grows faster than the $\propto \alpha^{1/7}$ scaling, with this non-power-law term manifesting at polarizabilities between those of Kr and Xe. 

Figure \ref{fig:polarize}(c-d) compares empirical scaling relations with classical force field parameters for He and Rn (red markers) with uncertainty quantification estimation (gray credibility interval). Credibility intervals were calculated by real unit scaling of the two standard deviation estimates on the reduced LJ units ($\pm 0.02 \sigma$ and $\pm 0.1 \epsilon$) reported in our recent study on classical force field uncertainty propagated from noisy scattering data \cite{shanks_bayesian_2024}. This gives upper and lower credibility interval bounds for the power law scaling relations of $\sigma_{int}:[3.097\alpha^{1/7}, 3.223\alpha^{1/7}]$ and $\epsilon_{int}:[0.135\alpha^{0.865}, 0.165\alpha^{0.865}]$. Note that these uncertainty estimates assume that the scattering data has a baseline noise level of 0.005 or less out to 30 $\si{\angstrom}^{-1}$. Error bars for Rn were reported according to Mick \textit{et al.} \cite{mick_prediction_2016}, but we speculate that the true uncertainty is larger due to the multimodal nature of their likelihood function response surface.

Table \ref{tab:ff} presents noble gas $\sigma$ and $\epsilon$ parameters estimated using SOPR-derived scaling relations, compared with existing force fields. Percent error (\%) is calculated in the usual way, $100(x_{pred} - x_{true})/x_{true}$, where $x_{pred}$ is the SOPR estimated parameter and $x_{true}$ is the independently optimized parameter. The estimates show errors below 3.9\% in all cases except for He's dispersion energy with a difference of 89\% from previously reported parameters \cite{tchouar_thermodynamic_2003}. This large error for He is likely due to strong nuclear quantum effects present in sub-critical liquid helium \cite{van_westen_corresponding-states_2025}. Speculative estimates for the parameters of liquid-phase Og are also included, though they cannot be validated by thermodynamic predictions since Og can only be synthesized one atom at a time \cite{smits_quest_2024}. Furthermore, accurate modeling of Og’s interatomic forces requires considering relativistic effects via the Dirac equation \cite{smits_oganesson_2020}, suggesting its properties may differ significantly from other noble gases. Despite these challenges, the parameters presented offer a theoretical foundation for investigating liquid-phase Og until experimental data becomes available. 

\begin{table}[H]
\centering
\caption{\label{tab:ff} Estimated $\sigma$ and $\epsilon$ parameters for the noble gas series using the QDO scaling relations. Deviations from previously reported force field parameters are given in percent error.}
\begin{tabular}{l c c c c c c c c}
\hline
\textrm{Element}&
\textrm{$\sigma$ (SOPR)}&
\textrm{$\sigma$}&
\textrm{Error (\%)}&
\textrm{$\epsilon$ (SOPR)}&
\textrm{$\epsilon$}&
\textrm{Error (\%)}&
\textrm{$\alpha$ ($\si\angstrom{}^3$)}\\
\hline
He     & 2.533 & 2.524 \cite{tchouar_thermodynamic_2003} & \textcolor{green}{0.4}  & 0.039 & 0.020 \cite{tchouar_thermodynamic_2003} & \textcolor{red}{89}     & 0.208 \cite{gaiser_polarizability_2018} \\
Ne     & 2.768 & 2.794 \cite{mick_optimized_2015}        & \textcolor{green}{-0.9} & 0.065 & 0.064 \cite{mick_optimized_2015}        & \textcolor{green}{1.5}  & 0.396 \cite{gobre_efficient_2016}       \\
Ar     & 3.393 & 3.405 \cite{mick_optimized_2015}        & \textcolor{green}{-0.4} & 0.242 & 0.242 \cite{mick_optimized_2015}        & \textcolor{green}{-0.1} & 1.661 \cite{gobre_efficient_2016}       \\
Kr     & 3.600 & 3.645 \cite{mick_optimized_2015}        & \textcolor{green}{-1.2} & 0.342 & 0.350 \cite{mick_optimized_2015}        & \textcolor{green}{-2.3} & 2.528 \cite{gobre_efficient_2016}       \\
Xe     & 3.858 & 3.964 \cite{mick_optimized_2015}        & \textcolor{green}{-2.7} & 0.469 & 0.484 \cite{mick_optimized_2015}        & \textcolor{green}{-3.1} & 4.118 \cite{gobre_efficient_2016}       \\
Rn     & 3.986 & 4.145 \cite{mick_prediction_2016}       & \textcolor{green}{-3.8} & 0.600 & 0.580 \cite{mick_prediction_2016}       & \textcolor{green}{3.4}  & 4.965 \cite{gobre_efficient_2016}       \\
Og     & 4.297 &                 -                       &            -            & 0.964 &                   -                     &           -             & 8.590 \cite{smits_oganesson_2020}       \\
\hline
\end{tabular}
\end{table}

Our results reveal a clear connection between structure-derived potentials and fundamental quantum mechanical theories of interatomic forces and electronic polarization, challenging the traditional view that structure-derived potentials are merely empirical representations of interatomic interactions \cite{toth_interactions_2007}. Furthermore, structure based methods like SOPR typically require only 10-20 classical MD simulations to learn an effective pair potential, providing a low computational cost alternative to expensive quantum mechanical calculations while being learned directly from experimental observation. Furthermore, SOPR potentials may provide an experimentally validated way to refine classical potential representations within corresponding-states frameworks for classical and quantum fluids \cite{van_westen_corresponding-states_2025}.  

Looking ahead, the physical insights gained from scattering analysis could play a key role in the design of biomolecular FFs, where standardized methods and benchmarks for interatomic interactions remain underdeveloped \cite{van_der_spoel_systematic_2021}. Developing FFs that can efficiently account for electronic polarization continues to be an active area of research, with significant implications for large-scale simulations of complex systems like membranes and protein complexes. Our findings indicate that structure refinement approaches may be well-suited for learning effective pair interactions consistent with polarization effects in liquid phases, potentially in less time than it typically takes to train empirical force fields (which often require hundreds or thousands of training simulations and the use of machine learning surrogate models to navigate the parameter space). Lastly, scaling relations between the $\sigma$ and $\epsilon$ parameters of structure-based potentials and atomic dipole polarization could provide valuable initial guesses for FF optimization or serve as a \textit{post hoc} validation of physical consistency with quantum theory.

\section{Conclusions}

In summary, collision and hard-particle diameters inferred from neutron scattering experiments in noble liquids were found to follow a quantum Drude oscillator scaling law with atomic dipole polarizabilities. This finding represents the first experimental evidence of QDO-type behavior in the liquid state. Additionally, the classical description of noble gas interatomic forces was shown to depend essentially on a single parameter: the atomic dipole polarizability. Empirical scaling laws on the classical $\sigma$ and $\epsilon$ parameters derived from SOPR potentials were then shown to closely reproduce state-of-the-art classical models of noble gases ranging from helium to radon. 

More broadly, the methodology presented in this work has the potential to advance research on inferring interatomic potentials from structural data. For many years, structure inversion techniques have been regarded as unable to yield physically accurate representations of interatomic interactions. Additionally, the focus on many-body and quantum mechanical methods has caused interest in developing approaches for investigating classical pairwise representations of interatomic forces to decline. Nevertheless, this study demonstrates that nonparametric inference of interaction potentials from high-resolution neutron scattering data can reveal subtle and complex aspects of intermolecular forces. It also highlights why machine learning interpretations of scattering experiments are instrumental in this respect.

\section*{Data Availability}

The datasets generated during and/or analyzed during the current study are available from the corresponding author on reasonable request.

\section*{Acknowledgments}

This study is supported by the EFRC-MUSE, an Energy Frontier Research Center funded by the U.S. Department of Energy, Office of Science, Basic Energy Sciences under Award No. DE-SC0019285.

\section*{Author Contributions}

B.L.S: conceptualization, algorithm development and implementation, data analysis/visualization, and writing. H.W.S: algorithm implementation. P. Jungwirth:  conceptualization, writing. M.P.H: conceptualization, funding acquisition.

\section*{Competing Interest}

The authors declare no conflicts or competing interests.

\section{Appendix}
\appendix
\input{si.tex}

\printbibliography

\end{document}

%% file: si.tex
\section{Structure Optimized Potential Refinement Settings and Molecular Dynamics}

The molecular simulation corrector is a Canonical ($NVT$) molecular dynamics (MD) simulation performed in HOOMD-Blue \cite{anderson_hoomd-blue_2020}. MD simulations were initiated with $N = 2916$ atoms on a cubic lattice of initial length of $3*N^{1/3}$ $\si\angstrom{}$ and performed in the NVT ensemble using HOOMD's native ConstantVolume integrator and Nose-Hoover thermostat with a default coupling constant of $\tau = 100 \delta t$. The timestep, $\delta t$, was chosen to be 0.5 femtoseconds and pair potential interactions were truncated at 3$\sigma$ with analytical tail corrections. MD simulations proceeded in four steps (1) thermalization of particle momenta for 5000 timesteps, (2) box size ramping to the experimental density using the hoomd.variant.Ramp command for 10,000 timesteps, (3) a 0.5 ns equilibration at the experimental density, and (4) a 0.5 ns production run. Radial distribution functions were calculated with Freud \cite{ramasubramani_freud_2020} sampled at $100$ timestep intervals. Convergence is checked against the average squared error between the simulated and experimental radial distribution function such that $\langle [\Delta g^{(n)}(r)]^2 \rangle < 5*10^{-4}$, which generally took 15-20 iterations at SOPR scaling constant $\gamma = 0.2$. 

\begin{table}[H]
\centering
\caption{\label{tab:refs}
Reduced temperature ($T_r = T/T_c$) and atomic density ($\rho$) are listed for the neutron scattering experimental conditions. 12-6 Lennard-Jones potentials with parameters ($\sigma_{ai}$,$\epsilon_{ai}$) were used as reference potentials.
}
\begin{tabular}{l c c c c c}
\hline
\textrm{Element}&
\textrm{$T_r$}&
\textrm{$\rho$ (1/$\si{\angstrom}^{3}$)}&
\textrm{$\sigma_{ai}$ ($\si{\angstrom}$)}&
\textrm{$\epsilon_{ai}$ (kcal/mol)}\\
\hline
Ne & 0.95 & 0.02477  & 2.76 & 0.122\\
Ar & 0.56 & 0.02125  & 3.35 & 0.287\\
Kr & 0.95 & 0.01187  & 3.58 & 0.582\\
Xe & 0.95 & 0.00881  & 3.89 & 0.811\\
\hline
\end{tabular}
\end{table}

The Gaussian process regression step used mean zero, $\mu = 0$, and a squared-exponential kernel with width parameter, $w = 0.05$ kcal/mol and length-scale parameter, $\ell = 1$ $\si\angstrom{}$. GPR regression was initiated at 0.9$\sigma$ which is well within the region of the potential that influences the MD. Experimental neutron scattering data and SOPR reference potentals are the same as the SOPR method paper \cite{shanks_transferable_2022} and reproduced below in Table \ref{tab:refs}. Source code for SOPR is available on GitHub \href{https://github.com/hoepfnergroup/SOPR}{here}.